\documentstyle[twocolumn,aps,psfig]{revtex}
\begin{document}
\title{Holographic principle and the dominant energy condition for Kasner type
metrics.}
\author{Mauricio Cataldo$^{\,\,a}$ {\thanks{%
E-mail address: mcataldo@ubiobio.cl}}, Norman Cruz$^{\,\,b}$
{\thanks{%
E-mail address: ncruz@lauca.usach.cl}}, Sergio del Campo $^{c}$ \thanks{%
E-mail address: sdelcamp@ucv.cl} and Samuel Lepe$^{b,c}$ {\thanks{%
E-mail address: slepe@lauca.usach.cl}}}
\address{$^a$ Departamento de F\'\i sica, Facultad de Ciencias, Universidad
del B\'\i o-B\'\i o, Avda. Collao 1202, Casilla 5-C, Concepci\'on,
Chile. \\ $^b$ Departamento de F\'\i sica, Facultad de Ciencia,
Universidad de Santiago de Chile, Avda. Ecuador 3493, Santiago,
Chile.\\ $^c$ Instituto de F\'\i sica, Facultad de Ciencias
B\'asicas y Matem\'aticas, Universidad Cat\'olica de Valpara\'\i
so, Avenida Brasil 2950,  Valpara\'\i so, Chile.} \maketitle

\begin{abstract}
{\bf {Abstract:}} In this letter we study adiabatic anisotropic
matter filled Bianchi type I models of the Kasner form together
with the cosmological holographic bound. We find that the dominant
energy condition and the holographic bound give precisely the same
constraint on the scale factor parameters that appear in the
metric.

\vspace{0.5cm}

PACS number(s): {98.80.Hw, 98.80.Bp}
\end{abstract}

\smallskip\
Motivated by the example of black hole entropy, recently a new set
of conjectures were put forward, which are known as ``the
holographic principle"~\cite{Hooft,Susskind1}. According to this
principle, under certain conditions all the information about a
physical system is coded on its boundary, implying that the
entropy of a system cannot exceed its boundary area in Plank
units. In other words, the holographic principle (HP) requires
that the degrees of freedom of a spatial region reside on the
surface of the region and the number of degrees of freedom per
unit area to be no greater than 1 per Planck area.

The main aim of the HP is to extend this conjecture to a broader
class of situations. In this direction, it is believed that the HP
must eventually have implications for cosmology. The first
interesting specific attempt to apply the HP to homogeneous
cosmological models was made by Fischler and Susskind
(F-S)~\cite{Fischler}. They showed that, if one tries to apply
this principle to any box of coordinate size $\Delta x^i \propto
R$ in flat space, there appear several problems. If one supposes
for the late universe a constant entropy density in comoving
coordinates, the entropy in that box is proportional to its
coordinate volume $R^3$, while the surface area of the box grows
like $[a(t)R]^2$. When we take the limit $R \longrightarrow
\infty$, the entropy always is larger than the area and the
principle is violated~\cite{Fischler}. Thus, they proposed to
compare not the area of any region, but namely the area of a
region of the size of the particle horizon $R_{_{H}}$, which is a
causally connected part of the universe, with the entropy of the
matter inside this region. In other words, the cosmological
formulation of the HP, due to the authors of~\cite{Fischler},
implies that the entropy of matter $S$ inside
the particle horizon must be smaller than the area $A$ of the horizon, i.e. $%
S/A < 1$. Their formulation was very successful for a wide class
of flat and open universes, but it did not apply to closed
universes. Nevertheless, other studies found that the HP,
according to the F-S-formulation, is satisfied for a closed
universe filled with two fluids, where one of them has an equation
of state, $P= \gamma \rho$, with $\gamma < -1/3$~\cite{Rama}, in
which $\rho$ and $P$ are the energy density and the isotropic
pressure respectively. This type of equation of state has been
considered in the so-called ``quintessence" models (QCDM) which
invoke such fluids in order to explain why our universe is
accelerating~\cite{Ratra}. Other modifications of the F-S
conjecture of the HP have been raised
subsequently~\cite{Easther,Linde,Bak,Veneziano,Wang}. It is
interesting to note that in~\cite{Wang} the authors proposed a HP
for inhomogeneous universes.

A remarkable conclusion of the F-S conjecture is that the HP is
valid for flat or open Friedmann-Robertson-Walker
(FRW)cosmological models with equation of state satisfying the
condition $0 \leq P \leq \rho$. It is interesting to remark that
this condition satisfies the Dominant Energy
Condition (DEC)~\cite{Hawking}, which can be written as $\rho \geq 0$ and $%
-\rho \leq P \leq \rho$. The latter condition is related to the
physical constraint that a sound wave cannot propagate faster than
light. In this direction recently Bousso provided a broader
formulation for HP. He conjectured the entropy $S$ crossing a
certain light-like hypersurface is bounded by a two-dimensional
spatial surface with area $A$. To protect his conjecture against
pathologies such as superluminal entropy flow, Bousso required
matter fields to satisfy the DEC. The author claim that his
conjecture is a universal law which is valid for all type of
space-times that satisfy Einstein's equations and DEC, therefore
in particular for cosmological solutions~\cite{Bousso}. However,
Lowe~\cite{Lowe} has pointed out a number of difficulties with the
Bousso's conjecture.

In this letter we analyze anisotropic universes described by a
Kasner type metric for checking the Bousso's conjecture and
establishing a relation between HP and DEC. 
In this respect Bousso has given many examples showing that the
holographic bound is satisfied for homogeneous but isotropic
cosmological models, including the closed FRW
universe~\cite{Bousso}.

At early times the presence of anisotropy is a very natural idea
to explore. Even though the universe, on a large scale, seems
homogeneous and isotropic at the present time, there are no
observational data that guarantee the isotropy in an era prior to
the recombination. In fact, it is possible to begin with an
anisotropic universe which isotropizes during its evolution by the
damping of this anisotropy via a mechanism of viscous dissipation.
The anisotropies described above have many possible sources: they
could be associated with cosmological magnetic or electric fields,
long-wavelength gravitational waves, Yang-Mills fields, axion
fields in low-energy string theory or topological defects such as
cosmic strings or domain walls, among others (see
ref.~\cite{Barrow} and references therein). The HP may restrict
the kind of the matter contents, which affect the geometry and
evolution of the universe, in view that HP envolves the particle
entropy of the universe. Then the universe itself conformed with
the HP may belong to a restrict class~\cite{Bak}.

In the following we consider a Bianchi type I metric of the Kasner
form
\begin{eqnarray}  \label{Metric}
ds^2= - dt^2 + t^{2 p_1} dx^2+ t^{2 p_2} dy^2+ t^{2 p_3} dz^2
\end{eqnarray}
where $p_1$, $p_2$ and $p_3$ are three parameters that we shall
require to be constants. Then expansion factors $t^{p_1}$,
$t^{p_2}$ and $t^{p_3}$ would be determined via Einstein's field
equations. The space is anisotropic if at least two of the three
$p_i$ ($i=1,2,3$) are different. The Kasner universe, in the
classical sense, refers to a vacuum cosmology, for which is
satisfied the constraints $p_1+p_2+p_3=1$,
$p^2_1+p^2_2+p^2_3=1$~\cite {MiThWe}. For matter filled universes
these constraints will no longer be true.

For simplicity of further calculations we introduce the symbols
$s$ and $q$ defined as
\begin{eqnarray}
s=p_1+p_2+p_3 \,\,\,\,\ and \,\,\,\,\, q=p^2_1+p^2_2+p^2_3.
\end{eqnarray}

For a Kasner type metric we have~\cite{Fischler}
\begin{eqnarray}  \label{cota}
S/A= \sigma \, \Pi_i R_{_{H,i}}/ \left [\Pi_{_j} t^{p_j}t^{1-p_j}
\right ]^{2/3},
\end{eqnarray}
where $\sigma$ is the entropy density in comoving coordinates,
which is supposed to be a constant, and $R_{_{H,i}}=t^{1-p_i}$
(with $i,j=1,2,3$) is the coordinate size of the particle horizon
in direction $i$. The denominator in equation~(\ref{cota}) is the
proper area of the horizon. Finally we obtain that $S/A =\sigma
t^{1-s}$. We write this expression in the form
\begin{eqnarray}  \label{cota holografica del Kasner}
S/A = \sigma \left( \frac{t}{t_p} \right)^{1-s}.
\end{eqnarray}
Here we consider that for $t \geq t_p$ the HP is valid, where
$t_p$ is the Planck time. So if the holographic bound $S/A\leq~1$
was satisfied at the Planck time, later on it will be satisfied
even better if
\begin{eqnarray}  \label{conditionHP}
1-s \leq0.
\end{eqnarray}

On the other hand, we shall require the model to satisfy the
dominant energy conditions specified by $- \rho \leq P_j \leq
\rho$~\cite{Hawking} where $\rho$ is the energy density and $P_j$
(with $j = x,y,z$) are the effective momenta related to the
corresponding coordinate axes. The DEC represents a very
reasonable physical condition for any gravitational field and
holds for all known forms of matter fields. This condition may be
interpreted as saying that to any observer the local energy
density appears non-negative and the local energy flow vector is
non-spacelike.

Let us write the dominant energy condition explicitly in terms of
the parameters that enter into the metric~(\ref{Metric}), i.e.
$p_1, p_2$ and $p_3$. In other words, we shall consider the matter
content in a general sense without any specification about the
kind of matter which fills the universe.

From the metric (\ref{Metric}) the Einstein field equations can be
written in comoving coordinates as
\begin{eqnarray}  \label{ro}
\frac{p_1 p_2 + p_1 p_3 + p_2 p_3}{t^2}=8 \pi G \, \rho,
\end{eqnarray}
\begin{eqnarray}  \label{P1}
-\frac{p_2^2 + p_3^2 -p_2 - p_3 + p_2 p_3}{t^2}=8 \pi G \, P_x,
\end{eqnarray}
\begin{eqnarray}  \label{P2}
-\frac{p_1^2 + p_3^2 -p_1 - p_3 + p_1 p_3}{t^2}= 8 \pi G \ P_y
\end{eqnarray}
and
\begin{eqnarray}  \label{P3}
-\frac{p_1^2 + p_2^2 -p_1 - p_2 + p_1 p_2}{t^2}= 8 \pi G \, P_z.
\end{eqnarray}
Note that both $\rho$ and $P_j $ (with $j=x,y,z$) scale as
$t^{-2}$. Thus, the dominant energy conditions will give some
specific relations among the constant Kasner parameters $p_i$.

The conditions $P_j \leq \rho$, with $j = x,y,z$, yield three
inequalities given by
\begin{eqnarray}  \label{ine1}
(s-p_i)(s -1)\geq 0,
\end{eqnarray}
where $i=1,2,3$ which, after adding them, reduce to just one
inequality given by
\begin{eqnarray}  \label{expresion 5}
2 s (s-1) \geq 0.
\end{eqnarray}
In a similar way, from $P_j \geq - \rho$ we get the inequalities
\begin{eqnarray}
s(1+p_i)-p_i \geq q,
\end{eqnarray}
(no sum over $i$). After adding them we get
\begin{eqnarray}  \label{expresion7}
s \geq \frac{3q-s^2}{2}.
\end{eqnarray}

It is easy to show that $3q-s^2=(p_1-p_2)^2+(p_1-p_3)^2+(p_2-p_3)^2$. Then $%
3q-s^2 \geq 0$ and from expression~(\ref{expresion7}) we conclude
that $s \geq 0$. With this condition on $s$, we obtain from
expression~(\ref {expresion 5}) that necessarily $1-s \leq 0$,
which is the same condition expressed by~(\ref{conditionHP}).
Thus, for the metric~(\ref{Metric}), the holographic
bound~(\ref{cota holografica del Kasner}) is always satisfied for
$t \geq t_p$. In other words we have shown that, for an
anisotropic Bianchi type I universe of Kasner form, the dominant
energy condition gives the same requirement as that obtained from
the holographic bound~(\ref {conditionHP}). We conclude that, for
Kasner metrics, if DEC is not satisfied then HP is not satisfied
at all and viceversa.

In general in cosmology the entropy is associated with the
particle species present in the expanding universe. Thus for the
classical vacuum Kasner space-time ($s=q=1$) we can not define an
entropy. For universes filled with any matter content, not
necessarily an ideal fluid ($P=\gamma \rho$ and DEC implies that
$-1 \leq \gamma \leq 1$), one of the following conditions is
satisfied: $s \neq 1$, $q=1$; $s=1$, $q \neq 1$ or $s \neq 1$, $q
\neq 1$. Nevertheless, the only ideal fluid that is compatible
with the anisotropic Kasner metric is the Zel'dovich
fluid~\cite{Brevik,Cataldo} which has the equation of state of
stiff matter ($\gamma=1$) given by
\begin{eqnarray}
\rho=P=\frac{1}{16 \pi t^2} (1-q)
\end{eqnarray}
for which $s=1$ holds and $q \neq 1$. In this case $S/A$ is
constant in time. Thus, depending on the boundary condition, the
HP may be saturated by this kind of universe.

For any ideal fluid, with $\gamma \neq 1$, we always obtain an
isotropic flat universe~\cite{Brevik}. In this case all constant
parameters $p_i=p$ ($ i=1,2,3$). Then $1-s=1-3p$ and the
holographic bound is valid if $3p\geq 1$.
It is easy to show that in this case the Einstein equations imply that $%
p=2/(3(\gamma +1))$. Then the holographic bound implies that
\begin{eqnarray}
\frac 2{1+\gamma }\geq 1.
\end{eqnarray}
This condition is always satisfied since $-1<\gamma \leq 1$ and
the holographic bound is satisfied for all $\gamma $ in the range
implied by the DEC. This fact generalizes the result obtained
in~\cite{Fischler}.

For isotropic flat cosmological models the same result was
obtained by Kaloper and Linde in Ref.~\cite{Linde}. In this
reference the authors argued that when the F-S conjecture is
applied to the inflationary expansion of the universe ($-1\leq
\gamma \leq -1/3$ for an ideal fluid), the evolution of the
universe should be considered not at $t=t_p$ but after reheating.
This follows from the fact that the density of matter after
inflation becomes negligibly small, so it must be created again in
the process of reheating of
the universe and this process is strongly non-adiabatic. This implies that $%
\sigma $ is no more a constant. In the case of the Kasner type
metric we can not include a quantum vacuum $\left( P=-\rho
=-\Lambda \right)$, i.e. $\gamma =-1$,  because for this case we
have an equation of state for which $(\gamma \neq 1)$ and, as we
mentioned above, the Kasner type metric isotropizes. However, we
can consider in principle the dissipative processes in this type
of metrics including a viscous fluid. In this case,  for an
anisotropic viscous fluid we have the energy-momentum tensor
\begin{eqnarray}
T_{\alpha \beta }=[\rho +(p-\xi \theta )]u_\alpha u_\beta -(p-\xi
\theta )g_{\alpha \beta }+2\eta \sigma _{\alpha \beta },
\label{tensor con shear}
\end{eqnarray}
where $u_\alpha $, $\rho $, $p$, $\xi $ and $\eta $ are the
fluid's four velocity, the energy density, the isotropic pressure,
the bulk and shear coefficients of viscosity respectively. It can
be shown,  at early times,
that the shear viscosity  is much greater than the bulk viscosity, i.e. $%
\eta >>\xi $~\cite{Caderni}. This means that the generation of
entropy is proportional to $\eta $ and therefore we have
non-adiabatic expansion of the universe.

For the Kasner type metric~(\ref{Metric}) it can be shown that the
shear viscosity has the form~\cite{Brevik}
\begin{eqnarray}
\eta= \frac{1}{16 \pi G}(1-s).
\end{eqnarray}
The second law of thermodynamics imposes that $\eta \geq 0$ and
then $1-s \geq 0$~\cite{Cataldo1,Brevik}. If we contrast this
condition with that obtained from HP and DEC, we conclude that
only $S=1$ satisfies all three conditions. Then the anisotropic
caracter of the metric is preserved. We could understand this as
follows, either HP may not be applied to non-adiabatic
processes~\cite{Linde} or the thermodynamics involved in the
problem is not appropriated due to his non-causal
structure~\cite{Maartens}. We hope to come back to this study in a
near future.


In conclusion we have shown that for an anisotropic Bianchi type I
universe of Kasner form the dominant energy condition gives the
same requirement as that obtained from the holographic bound when
the processes involved are adiabatic. Perhaps there exists in
general a deep relationship between the DEC and the holographic
bound, when both are applied to adiabatic flat cosmology.

The results obtained here can be applied for Kasner type metrics
in any dimensional $n+1$ space-time.

\mbox{} \\ \mbox{} \mbox{} \mbox{}

We thank Paul Minning for carefully reading the manuscript. NC and
SL acknowledge the hospitality of the Physics Department of
Universidad del B\'\i o-B\'\i o and all authors acknowledge the
hospitality of the Physics Department of Universidad de la
Frontera where this work was partially done. This work was
supported by COMISION NACIONAL DE CIENCIAS Y TECNOLOGIA through
Grants FONDECYT N$^0$ 1010485 (MC and SdC), N$^0$ 1000305 (SdC)
and N$^0$ 2990037(SL). Also MC was supported by Direcci\'on de
Promoci\'on y Desarrollo de la Universidad del B\'\i o-B\'\i o,
SdC was supported from UCV-DGIP (2001) and NC was supported by
USACH-DICYT under Grant N$^0$ 04-0031CM.

\end{document}